\documentclass[sigconf]{acmart}

\usepackage[T1]{fontenc}

\usepackage[]{xcolor}
\newcommand{\TODO}[1]{\textcolor{red}{#1}\GenericWarning{}{LaTeX Warning: TODO: #1}}\newcommand\todo\TODO

\usepackage[nolist]{acronym}
\usepackage{paralist}
\usepackage{hyperref}
\usepackage{cleveref}
\usepackage{graphics}
\usepackage{enumitem}
\crefname{lstlisting}{listing}{listings}
\Crefname{lstlisting}{Listing}{Listings}
\usepackage{listings}
\usepackage[numbered]{mcode}
\lstdefinelanguage{diff}{
  morecomment=[f][\color{blue}]{@@},     %
  morecomment=[f][\color{red!60!black}]-,         %
  morecomment=[f][\color{green!60!black}]+,       %
  morecomment=[f][\color{magenta}]{---}, %
  morecomment=[f][\color{magenta}]{+++},
}

\usepackage{graphicx}
\usepackage[hang,small,bf,tight]{subfigure}
\usepackage{svg}
\usepackage{tcolorbox}
\usepackage{pifont}
\usepackage{array}
\usepackage{booktabs}
\usepackage{multirow}
\usepackage{colortbl}
\usepackage{spreadtab}
\usepackage{longtable}
\usepackage{pdflscape}
\usepackage{float}
\usepackage{multicol}

\usepackage{xspace}
\newcommand{\mufin}{\textsc{MUFIN}\xspace}
\newcommand{\selfapr}{\textsc{SelfAPR}\xspace}
\newcommand{\buglab}{\textsc{BugLab}\xspace}

\begin{document}

\title{MUFIN: Improving Neural Repair Models with Back-Translation}

\author{André Silva}
\affiliation{
  \institution{KTH Royal Institute of Technology}
  \country{Sweden}
}
\email{andreans@kth.se}

\author{João F. Ferreira}
\affiliation{
    \institution{INESC-ID and IST, University of Lisbon}
    \country{Portugal}
}
\email{joao@joaoff.com}

\author{He Ye}
\affiliation{
  \institution{KTH Royal Institute of Technology}
  \country{Sweden}
}
\email{heye@kth.se}

\author{Martin Monperrus}
\affiliation{
  \institution{KTH Royal Institute of Technology}
  \country{Sweden}
}
\email{monperrus@kth.se}

\begin{acronym}[H.264/SVC]
    \acro{AST}{Abstract Syntax Tree}
    \acro{APR}{Automated Program Repair}
    \acro{BIFI}{Break-It-Fix-It}
    \acro{DNN}{Deep Neural Network}
    \acrodefplural{DNN}{Deep Neural Networks}
    \acro{FL}{Fault Localization}
    \acro{LSTM}{Long Short-Term Memory}
    \acro{NMT}{Neural Machine Translation}
    \acro{NLP}{Natural Language Processing}
    \acro{RNN}{Recurrent Neural Network}
    \acrodefplural{RNN}{Recurrent Neural Networks}
    \acro{SSL}{Self-Supervised Learning}
\end{acronym}

\begin{abstract}
Automated program repair is the task of automatically repairing software bugs.
A promising direction in this field is self-supervised learning, a learning paradigm in which repair models are trained without commits representing pairs of bug/fix.
In self-supervised neural program repair, those bug/fix pairs are generated in some ways.
The main problem is to generate interesting and diverse pairs that maximize the effectiveness of training.  
As a contribution to this problem, we propose to use back-translation, a technique coming from neural machine translation.
We devise and implement \mufin, a back-translation training technique for program repair, with specifically designed code critics to select high-quality training samples.
Our results show that \mufin's back-translation loop generates valuable training samples in a fully automated, self-supervised manner, generating more than half-a-million pairs of bug/fix.
The code critic design is key because of a fundamental trade-off between how restrictive a critic is and how many samples are available for optimization during back-translation. 
\end{abstract}

\maketitle

\section{Introduction}

Over the last decades, software systems have evolved to become some of the most complex human artifacts ever.
Developing them is, typically, a multi-step effort realized by teams composed of differently skilled individuals.
Debugging, the activity of finding and fixing software bugs, is one of the most demanding activities of software development~\cite{vessey1985expertise}.
It requires developers to analyze and understand code and error logs, code they have not written themselves, and to find an edit to the program that fixes the bug, aka a patch.

\acf{APR}~\cite{goues2019automated} is an active research domain about automatically finding patches to software bugs without requiring human developer time.
APR is foreseen to save valuable person-hours, reduce the costs of developing and maintaining software systems, and decrease the number of people required in such endeavors, ultimately leading to an ever-automated world.

Traditional APR approaches (e.g., \cite{forrest2009genetic,kim2013automatic,liu2019tbar,cornu2015npefix,nguyen2013semfix,xuan2016nopol,mechtaev2016angelix}) either rely on manually defined rules for changing programs or are based on constraint inference and solving, narrowing their scope to only a portion of software bug types.
On the other hand, more modern learning-based APR approaches (e.g., \cite{gupta2017deepfix,chen2019sequencer,lutellier2020coconut,ye2021neural,chen2021neural,yasunaga2021break,allamanis2021self,yasunaga2020graph}) are agnostic to bug types and learn repair transformations, without manual intervention, from the training data. 

Learning-based \ac{APR} is data-hungry.
Previous work in supervised learning for \ac{APR} has made huge efforts to create datasets of bugs and their fixes, typically from commits in code repositories (e.g., GitHub, GitLab), student assignments, and code competitions.
While collecting code on disk scales, executing failures and fixes does not because of the absence of configuration files and execution environments.
Consequently, the datasets of executable code, put together, add up to just a few thousand samples \cite{just2014defects4j, madeiral2019bears, lin2017quixbugs, saha2018bugs}.

\acf{SSL} is a solution to the problem of collecting executable pairs of bug/fix \cite{ye2022selfapr}.
Instead of collecting past commits, the idea is to generate training samples, with an automated procedure to synthesize pairs of bug/fix that can be used to train a neural network.
Self-Supervised Learning is a relatively unexplored area of APR research with only a few papers on syntax errors \cite{yasunaga2020graph,yasunaga2021break}, simple Python errors \cite{allamanis2021self} and repair with execution diagnostics \cite{ye2022selfapr}.

In this paper, we propose \mufin, a novel self-supervised approach to functional program repair.
The main goal of \mufin is to radically improve the generalization capability of a given neural program repair model.
The key innovation of \mufin is the use of back-translation paired with a code critic that selects high-quality samples from the ones that are generated.
Back-translation always employs two models, in our case a fixer and a breaker. 
\mufin fine-tunes both the breaker and the fixer models alternately, using the fixer model to generate training samples for the breaker and vice-versa.
This training data generation strategy enables the model to be exposed to more and more diverse training data, improving generalization over the unseen testing dataset.
To the best of our knowledge, we are the first to apply this concept to functional program repair.

We evaluate \mufin on two widely-accepted benchmarks for program repair in Java: QuixBugs \cite{lin2017quixbugs} and Defects4J \cite{just2014defects4j}.
Our results show that \mufin out-performs the baseline models, correctly repairing +12 (Defects4J) and +4 (QuixBugs) bugs than the second best-performing model.
Furthermore, we study the impact of the critic design in \mufin.
Our experimental results confirm the importance of the critic choice by revealing a trade-off between critic restrictiveness and the quantity of training samples.
The best performing critic is based solely on the compilation results: more and less permissive critics achieve worse effectiveness.

To sum up, the main contributions of this paper are:
\begin{itemize}
    \item We design and implement \mufin: a novel self-supervised approach for functional program repair. As opposed to collecting commits for training, \mufin generates valuable training samples.
    
    \item We conceive three families of code critics for self-supervised functional program repair. To the best of our knowledge, we are the first to study back-translation critics for program repair in Java.

    \item We perform a set of experiments to measure to what extent \mufin improves over a strong baseline. Our results demonstrate the usefulness of \mufin in improving the quality of the generated patches and in improving patch compilability.

    \item We make the code, datasets, and experimental results of our study, publicly available at \url{https://github.com/andre15silva/mufin}.
\end{itemize}

\section{The MUFIN Approach}

\begin{figure*}[t]
    \centering
    \includegraphics[width=0.90\textwidth]{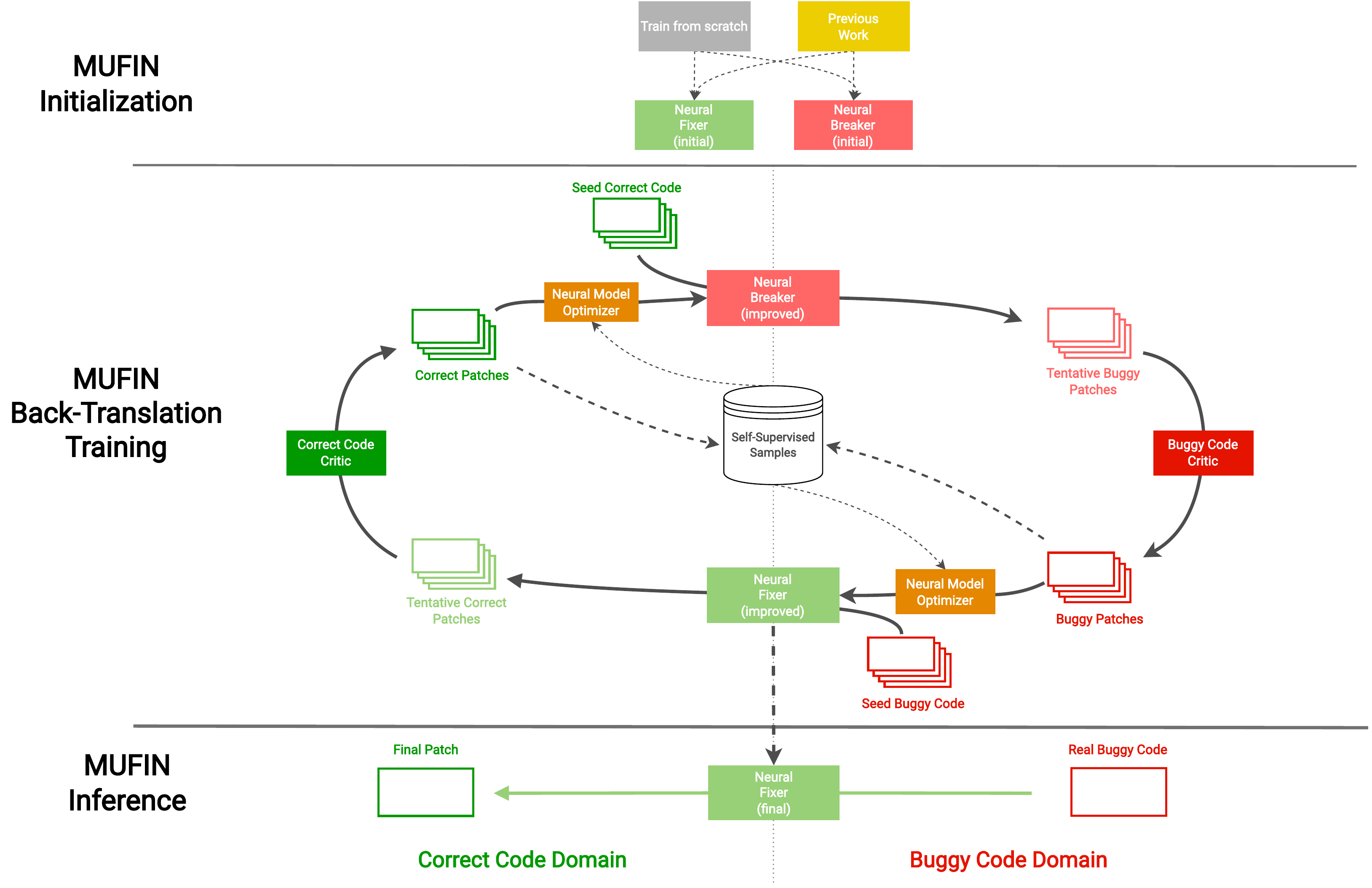}
    \caption{Overview of the \mufin approach.}
    \label{fig:mufin_overview}
\end{figure*}

\mufin is an original self-supervised approach for automated program repair.
It is illustrated in \Cref{fig:mufin_overview}.
The core novelty of \mufin is the training procedure which combines three training stages, the latter two being part of a loop.

\mufin requires an initialized neural breaker and neural fixer (\emph{MUFIN Initialization}).
The goal of the neural fixer is to generate correct code from buggy code, and that of the neural breaker is to generate buggy code from correct code.
Both can come from previous work if available or can be trained from scratch, see \autoref{sec:mufin_init}.

In the second stage, \mufin uses a back-translation loop (\emph{MUFIN Back-Translation}) to fine-tune both the breaker and the fixer models alternately.
Back-translation involves using a model that translates from one domain to another to generate new training samples for a second model that translates in the opposite direction, and vice versa, iteratively improving both models~\cite{lample2018phrase}.

The primary goal of \mufin Back-Translation is to accumulate valuable training samples for both the breaker and fixer in a completely self-supervised manner.
The generated training samples are stored and used collectively at each optimization step.
This results in an increasing number of training samples being used for fine-tuning with each iteration, improving the performance of the models over time.

The outcome of the entire training process is a high-quality neural fixer.
The automated generation of training samples is crucial in exposing the model to a diverse set of samples and thus improving its generalization ability.
The core advantage of the training process is that it is entirely self-supervised and does not require the tedious and expensive collection of pairs of bug/fix as done in supervised program repair.

We now describe each component of \mufin in detail.

\subsection{MUFIN Initialization}
\label{sec:mufin_init}

The \mufin Initialization stage serves the purpose of initializing two neural models: a neural fixer and a neural breaker.
Both models need to be available to bootstrap the subsequent \mufin Back-Translation stage.
They must exhibit reasonable performance, as their outputs are used as training samples during the back-translation loop.

\paragraph{Fixer Initialization} 
A neural fixer generates correct code from buggy code.
A reasonable neural fixer can be obtained through various means:
1) from past research having produced a publicly-available, reusable, and fine-tunable fixer model,
2) through traditional supervised training with commits,
or
3) by doing self-supervised training with mechanically-generated samples, with samples (i.e., from artificial buggy code to correct code).

\paragraph{Breaker Initialization}

A neural breaker generates buggy code from correct code.
While fixer models have been produced and shared in past research, breaker models  are very rare~\cite{tufano2019learning}.
Thus, there is an asymmetry between the availability of breakers versus fixers, calling specific work for initializing a breaker. As a matter of fact, training a breaker from scratch is, at the moment, necessary for implementing \mufin.

In \mufin Initialization, the training data used for initializing the breaker is generated by a mechanical breaker:
a mechanical breaker is composed of manually defined corruption rules that modify correct code such that it becomes buggy (it is not a neural network). A mutation testing tool is an example of such a mechanical breaker.

Given a mechanical-breaker, \mufin starts with correct programs which have a passing test suite.
For each correct program, the mechanical breaker generates multiple bugs by applying each corruption rule to multiple locations inside the correct sample.

\Cref{lst:mechanical_perturbation} gives an example of a bug generated by the mechanical breaker.
In this example, the mechanical breaker swaps the first two parameters of the method call \textit{serializeFields}.
By modifying the program like this, the mechanical breaker modifies its behavior and introduces a bug.

\begin{figure}[!h]
\begin{lstlisting}[language=diff,caption=Example of a training sample generated by a mechanical breaker (\selfapr~\cite{ye2022selfapr}). The mechanical breaker swaps the first two parameters of the method call \textit{serializeFields}\, modifying the intended behavior of the program.,numbers=none,label=lst:mechanical_perturbation]
--- a/correct.java
+++ b/buggy.java
@@ -552,11 +551,11 @@
  } else {
-  serializeFields(value, gen, provider);
+  serializeFields(gen, value, provider);
  }
\end{lstlisting}
\end{figure}

\subsection{MUFIN Back-Translation}
\label{sec:mufin_bt}

The goal of the \mufin Back-Translation stage is to improve the initialized fixer and breaker models by iteratively using one's output to train the other.
\mufin accomplishes this by using unpaired data (i.e., correct samples which are not linked to a buggy version or buggy samples which are not linked to a fixing patch).
A critic (\Cref{sec:critics}) filters outputs that do not meet a quality criterion, with the intent of maximizing the quality of the samples used for training.

The process begins by either applying the initial breaker to seed correct programs or by applying the initial neural fixer to seed buggy programs. Assuming the latter, the tentative correct patches output by the fixer are then filtered according to a correct code critic: a correct code critic filters samples and keeps only the ones it considers correct.
Then, the neural breaker is fine-tuned with the self-supervised samples to translate from correct programs to buggy programs. 

Given the improved neural breaker, tentative buggy patches are generated from correct programs.
In turn, the tentative buggy patches are filtered according to a buggy code critic:
a buggy code critic filters samples and keeps only the ones it considers buggy.
Then, the neural fixer is fine-tuned with the self-supervised samples to translate from buggy programs to correct programs.

\mufin Back-Translation can be configured to run for $N$ iterations and to generate $K$ tentative patches at each generation step using beam search.
In each iteration, both neural models are improved with back-propagation.
The data generated from one iteration is used in subsequent iterations as seed data.

At the end of the back-translation loop, one can throw away the breaker if we only do program repair, but the breaker may be reused in other tasks, see \autoref{sec:breakerrepurpose}.
The most important outcome is the final neural fixer, which is subsequently used for inference.

\subsection{Critics In Back-Translation}
\label{sec:critics}

In back-translation, the generation of high-quality training samples presents a significant challenge, as it can hamper the effectiveness of the process~\cite{yasunaga2021break, roziere2021leveraging}.
To address this issue, one employs ``critics'' to evaluate the quality of the generated samples.
Critics are functions that judge a sample based on a quality criterion (\autoref{def:critic}).
In the context of back-translation for program repair, one needs two critics: one for keeping high-quality samples for training the neural fixer, and another one for keeping high-quality samples for training the neural breaker.

\begin{definition}[Critic]
\label{def:critic}
A critic is a predicate function $\mathcal{C}: \mathcal{X} \rightarrow \{0,1\}$, where $\mathcal{X}$ is the space of programs, and $\mathcal{C}(x)=1$ if $x$ is deemed acceptable, and $\mathcal{C}(x)=0$ otherwise.
\end{definition}

Critics can be configured with varying levels of restrictiveness.
Increasing the restrictiveness ensures that only high-quality samples are selected, while reducing the number of kept training samples.
We now describe two different families of critics we consider in \mufin, with different degrees of restrictiveness:

\paragraph{Compiler-based critics}
A critic may be based on compilation alone.
If the compilation process fails, the program cannot be executed, and the presence of functional bugs cannot be evaluated.
The behavior of each critic is as follows:

\emph{\textit{compiler} correct code critic}: Keeps all programs that compile successfully.

\emph{\textit{compiler} buggy code critic}: Keeps all programs that compile successfully. The idea is that we want to network focus on functional bugs which compile but have failing test suites, and not on compiler bugs.

\paragraph{Tests-based critics}
A critic may be based on test execution.
Failing unit tests are existential proofs of the presence of a functional bug.
The behavior of each critic is as follows:

\emph{\textit{tests} correct code critic}: Keeps all programs that compile and pass all tests successfully.

\emph{\textit{tests} buggy code critic}: Keeps all programs that compile successfully but have at least one failing unit test.

\subsection{MUFIN Inference}

\mufin Inference corresponds to the application of \mufin on real-world bugs.
During this stage, \mufin utilizes the final neural fixer to repair the buggy code.
Once presented with the code to be fixed, per previous work \cite{chen2019sequencer}, \mufin employs fault localization (e.g., GZoltar~\cite{campos2012gzoltar}, Flacoco~\cite{silva2021flacoco}) to identify a list of suspicious locations.
The neural fixer is then applied to these locations.
For each location, \mufin uses beam search to generate and rank a list of $K$ patches.

\subsection{Breaking Location Selection}
\label{sec:breaking_location_selection}

In the back-translation loop, an important aspect is the selection of locations on which the breaker model will operate to generate bugs.
With \mufin handling projects consisting of multiple source code files, there is a challenge in choosing the right locations to corrupt.

One naive solution would be to randomly select source code lines.
However, randomly selecting source code lines may not be effective, as the chosen locations may not be relevant.
For example, a line outside the class declaration could be chosen, leading to additional noise being introduced in the back-translation process.

To tackle this challenge, \mufin iterates over each file of each program and uses AST-based analysis~\cite{pawlak2016spoon} to identify statements located inside code blocks.
For each statement, a pair of line numbers indicating the starting and ending line numbers of the statement are outputted.
Given the identified locations, the breaker model receives their input representations and returns a number of tentatively buggy snippets of code.

\subsection{Input Representation}

The neural models are presented with a sequence of tokens.
To create this input representation, we follow existing research (\cite{ye2022selfapr, chen2019sequencer, ye2021neural}) and incorporate contextual information and fault localization information about the code snippet that requires modification.
Our input representation comprises three parts that are concatenated.
The first part contains the $n$ lines (configurable) that precede the code section requiring modification.
The second part is delimited by two special tokens, \texttt{[START\_BUGGY]} and \texttt{[END\_BUGGY]}, which mark the beginning and end of the code portion to be modified.
Lastly, the third part includes the lines that follow the code portion to be modified.
The same representation is used for both the neural fixer and neural breaker models.

\subsection{Implementation}

We reuse He et al.'s mechanical breakers used in \selfapr \selfapr~\cite{ye2022selfapr}.
We implement \mufin's neural models with a state-of-the-art encoder-decoder Transformer architecture T5~\cite{raffel2020exploring}, from HuggingFace.
We set the model dimensions to follow \texttt{t5-small}, with: ${d}_{model}$ = 512, ${d}_{ff}$ = 2048, ${d}_{kv}$ = 64, 8-headed attention mechanisms and 6 layers in each the encoder and decoder for a total of approximately 60M parameters.
We use the PLBART~\cite{ahmad-etal-2021-unified} SentencePiece tokenizer from HuggingFace.
We hold out 2\% of the training data for validation.
The models are optimized by AdamW~\cite{loshchilov2018decoupled,kingma2014adam}, with a batch size of 16, a learning rate of 1e-4, and a gradient decay of 0.01, on a single GPU (NVIDIA RTX A4000).
At each training stage, we use an early-stopping loop with validation loss.

We set $K$ to 10 when generating correct code in \mufin Back-Translation, following previous work~\cite{yasunaga2021break}, and to 1 when generating buggy code due to the number of code locations available.
We set $K$ to 100 during \mufin Inference, a value within the range of related work~\cite{zhu2021syntax, ye2022selfapr, chen2019sequencer}.
We employ early stopping during patch generation, such that generation stops when all beam sequences have reached the EOS token.

\section{Experimental Methodology}
 
\subsection{Research Questions}

\begin{itemize}
    \item \textbf{RQ1 (Impact of Back-Translation)}: What is the impact of \mufin's back-translation compared with a baseline model without back-translation?
    \item \textbf{RQ2 (Impact of Critics)}: To what extent does the critic design impact \mufin's effectiveness?
\end{itemize}

\subsection{Dataset Construction}

To evaluate \mufin, we construct datasets for both training and testing purposes.
We constrain our training and testing datasets to single-hunk bugs, due to the limitations of current neural models in repairing multi-location bugs~\cite{saha2019harnessing}.

For training, we construct two input datasets from the original Bears dataset~\cite{madeiral2019bears}: one comprising correct programs and another composed of buggy programs.
Bears is a dataset of reproducible Java bugs.
It contains 251 bugs collected from 72 different open-source projects.
We choose Bears as its samples are accompanied by a compiler and test suite, which \mufin's critics require.

For testing, we consider two widely adopted benchmarks: Defects4J~\cite{just2014defects4j} and QuixBugs~\cite{lin2017quixbugs}.
Defects4J v2.0 contains 835 bugs collected from 17 different open-source projects.
QuixBugs contains 40 programs collected from the Quixey Challenge.

We now explain in detail how each dataset is constructed.

\subsubsection{Training Datasets}
\label{sec:train_dataset}

\paragraph{Correct Program Dataset}
We follow the approach of Ye et al.~\cite{ye2022selfapr} to extract the earliest sample of each project.
From each sample, we use the patched version as a correct program.
This ensures that we do not over-sample from any project, given each project is represented by several samples.
Programs that cannot be reproduced locally are filtered out to ensure data consistency.

From the original Bears dataset, we obtain a total of 56 correct programs.
The discrepancy between the number of projects and the number of correct programs is due to failures in reproducing 16 correct programs, out of which 14 fail to compile and 2 fail to pass the tests.
Overall, the dataset consists of 1,342,614 lines of code and 70,160 unit tests.

\paragraph{Buggy Program Dataset}
To construct the buggy program dataset, we attempt to reproduce all available samples locally and remove the ones which we fail to reproduce.
For each sample, we keep only the buggy version.
Our final collection consists of 61 buggy programs from the Bears dataset.

\subsubsection{Testing Datasets}

\noindent To create our testing datasets, we apply filters to the Defects4J and QuixBugs datasets, selecting only bugs that meet two criteria: (i) are reproducible, and (ii) are not included in the training datasets.

We collect a total of 428 test samples from Defects4J and 39 test samples from QuixBugs.
Notably, we exclude all bugs from the \textit{JacksonDatabind} project to prevent data spillover.
This decision is made because both Bears (used in training) and Defects4J (used in testing) contain samples collected from this project.
We have taken precautions to prevent further cross-dataset contamination and are not aware of any other such instances.

\subsection{Baseline Model}
\label{sec:baseline_model}

\mufin Back-Translation requires reasonable models as input: a fixer and a breaker.
For our experiments, we need to control for model configuration (i.e., training dataset, input representation, architecture, hyper-parameters).
To this end, we train neural fixers and breakers from scratch. This gives us a baseline model for which we can demonstrate that back-translation improves.

Our baseline models are trained with mechanically-generated samples produced by the corruption model of \selfapr~\cite{ye2022selfapr} applied to the correct program dataset.
The neural breaker is trained to translate correct programs to buggy programs, while the neural fixer is trained to translate buggy programs to correct programs.
These models are subsequently used to initialize \mufin Back-Translation in RQ1 (\autoref{sec:methodology_rq1}) and RQ2 (\autoref{sec:methodology_rq2}).
For giving more perspective, we also use a baseline model trained according to related work BugLab \cite{allamanis2021self}, named accordingly.

\subsection{Methodology for RQ1}
\label{sec:methodology_rq1}

In RQ1, we compare \mufin with the baseline model described in \autoref{sec:baseline_model}.
To this end, we pick the best \mufin model according to hyper-optimization on the configuration space, including the critic in \mufin Back-Translation.

In order to evaluate the performance of the neural fixer models, we use the widely accepted Defects4J~\cite{just2014defects4j} and QuixBugs~\cite{lin2017quixbugs} benchmarks.
We compute 1) the number of plausibly repaired bugs (i.e., when the human-written tests pass) and 2) the number of correctly repaired bugs (i.e., when the generated patch is equivalent to the human-written patch, checked manually).

To eliminate biases inherited from the fault localization step, we consider perfect fault localization in our experimental setup, as done in related work~\cite{chen2019sequencer,ye2022selfapr,lutellier2020coconut}.
This approach enables us to focus solely on the patch generation step of the \ac{APR} pipeline and facilitates a fair comparison of performance across different approaches~\cite{liu2019you}.

\subsection{Methodology for RQ2}
\label{sec:methodology_rq2}

In RQ2, we study the extent to which the critic design impacts the overall effectiveness of \mufin.
To this end, we fine-tune the baseline model described in \autoref{sec:baseline_model} with \mufin Back-Translation using three different strategies: 1) without a critic, 2) with the \textit{compiler} critic family, and 3) with the \textit{tests} critic family.
The critic families are described in \autoref{sec:critics}.
The models are trained and evaluated following the methodology of RQ1.
Additionally, we compute the percentage of compilable patches generated by each model on each test dataset to measure the syntactic quality of the generated patches.

\section{Experimental Results}

\subsection{RQ1 Results (Back-Translation)}

In RQ1, we compare \mufin's repair effectiveness with the baseline model described in \autoref{sec:baseline_model} and BugLab \cite{allamanis2021self}.
The results of the evaluation are presented in \Cref{tab:rq1_results}, which shows the performance of each approach on both test datasets.
The table is structured as follows:
the first column displays the approach used to train each model, while the second and third columns indicate the number of training samples used in each training stage, with column \# 3 being the number of generated samples with self-supervision in \mufin Back-Translation.
The last two columns give the testing results on both QuixBugs and Defects4J.
For each cell X/Y, X denotes the number of correctly repaired bugs (i.e., when the human-written tests pass), while Y indicates the number of plausibly repaired bugs (i.e., when the human-written tests pass).

\buglab \cite{allamanis2021self}, which is optimized with 821,311 samples during the \mufin Initialization stage, can repair 2 and 16 bugs from QuixBugs and Defects4J, respectively.
Additionally, it can plausibly repair 4 and 31 bugs from QuixBugs and Defects4J, respectively.

The baseline model, which is optimized with 3,942,935 samples during the \mufin Initialization stage, can repair 2 and 16 bugs from QuixBugs and Defects4J, respectively.
Additionally, it can plausibly repair 4 and 43 bugs from QuixBugs and Defects4J, respectively.

\begin{table}
\centering
\caption{\mufin's repair effectiveness w.r.t. state-of-the-part self-supervised functional program repair approaches. \mufin correctly and plausibly fixes more bugs than any other approach over two benchmarks.}
\label{tab:rq1_results}
\resizebox{\columnwidth}{!}{
\begin{tabular}{c|cc|cc}
\toprule
Approach & \multicolumn{2}{c|}{\# Training Samples} & \multicolumn{2}{c}{Testing} \\
   &               Initialization & Self-Supervised &   \shortstack{QuixBugs \\ (39 bugs)} & \shortstack{D4J \\ (428 bugs)} \\
\midrule
   \buglab~\cite{allamanis2021self} &  821,311   & -   & 2/4 & 16/31  \\
   Baseline Model    &  3,942,935 & -   & 2/4 & 16/43  \\
   \textbf{\mufin}          &  3,942,935 & 197,959 & \textbf{6/7} & \textbf{28/62}  \\
\bottomrule
\end{tabular}
}
\end{table}

\mufin's golden model correctly repairs 28 bugs from Defects4J and 6 bugs from Quixbugs.
\mufin clearly outperforms the baseline model:
\mufin correctly repairs +12 Defects4J bugs and the same trend is observed in QuixBugs.
Since \mufin is based on the exact same baseline model, it means that the improvement comes from \mufin's core contribution: the back-translation loop.
By iteratively improving the neural fixer and breaker models using one's output to train the other, \mufin generates valuable training samples which improve generalization over the unseen testing datasets.
Similarly, better performance is observed for plausible patches: \mufin plausibly repairs 62 bugs from Defects4J and 7 bugs from QuixBugs, which is better than the baseline.

\paragraph{Generalization Example}
Let us now consider a bug that only \mufin's model can repair.
For example, \Cref{lst:cli_5_patch} shows \mufin's correct patch for bug \textsc{Cli-5} from Defects4J.
In this patch, the right-hand-side value of an assignment statement is modified from an arithmetic operation to a literal.
Neither the baseline model nor \buglab are able to correctly repair this bug.
This suggests that \mufin generated valuable training samples modifying arithmetic expressions.

\begin{figure}[!h]
\begin{lstlisting}[language=diff,caption={\mufin's correct patch for bug \textsc{Cli-5} from Defects4J. Neither the baseline model nor \buglab models able to correctly repair this bug.},numbers=none,label=lst:cli_5_patch]
--- a/buggy.java
+++ b/correct.java
@@ -552,11 +551,11 @@
  // stops infinite loop happening
-  nextLineTabStop = width - 1;
+  nextLineTabStop = 1;
  }
\end{lstlisting}
\end{figure}

\begin{figure}[htb]
	\centering
	\subfigure[QuixBugs]{\label{fig:repair_quixbugs}
	    \includegraphics[width=0.325\textwidth]{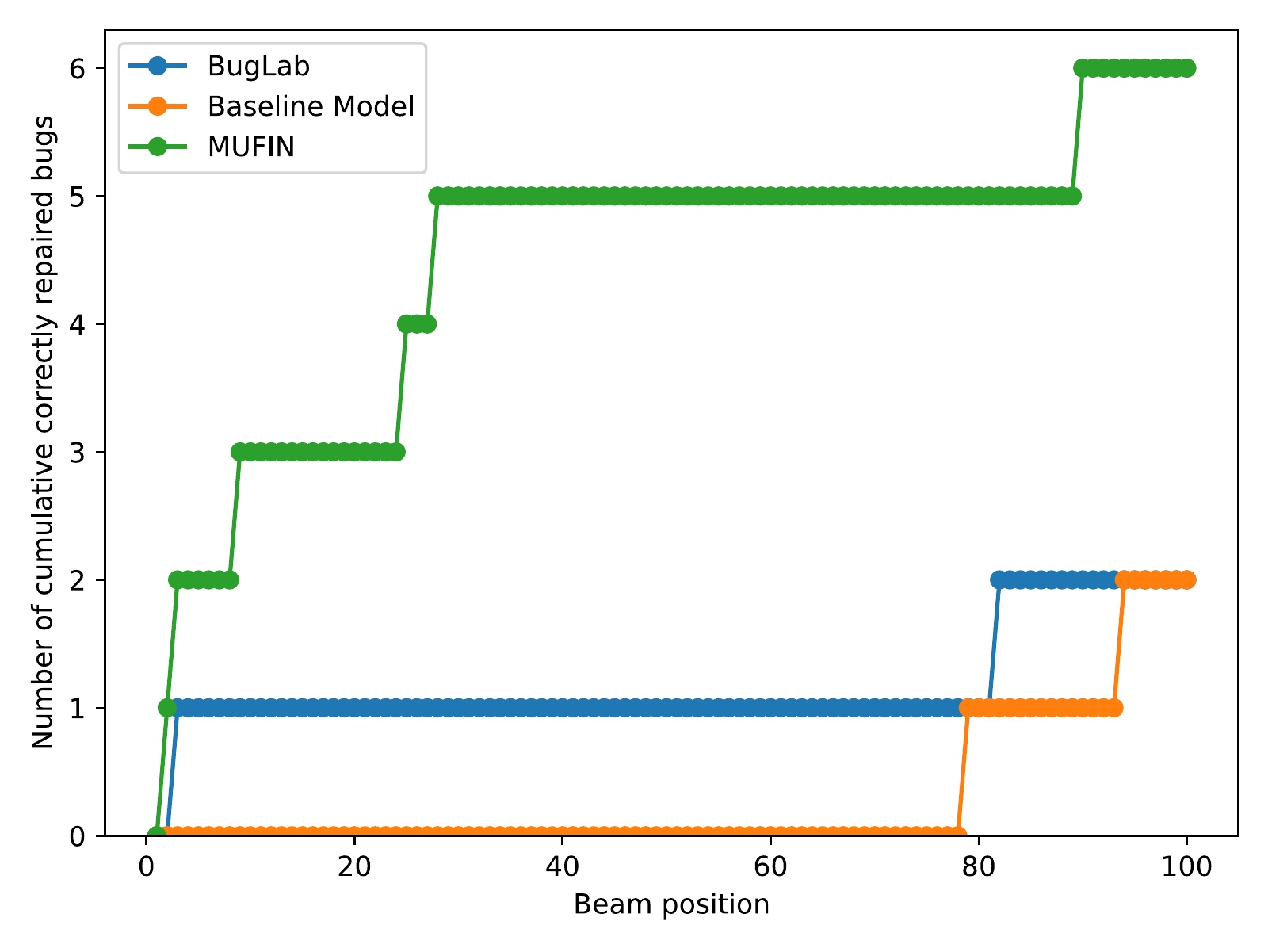}}
	    \qquad
	\subfigure[Defects4J]{\label{fig:repair_defects4j}
		\includegraphics[width=0.325\textwidth]{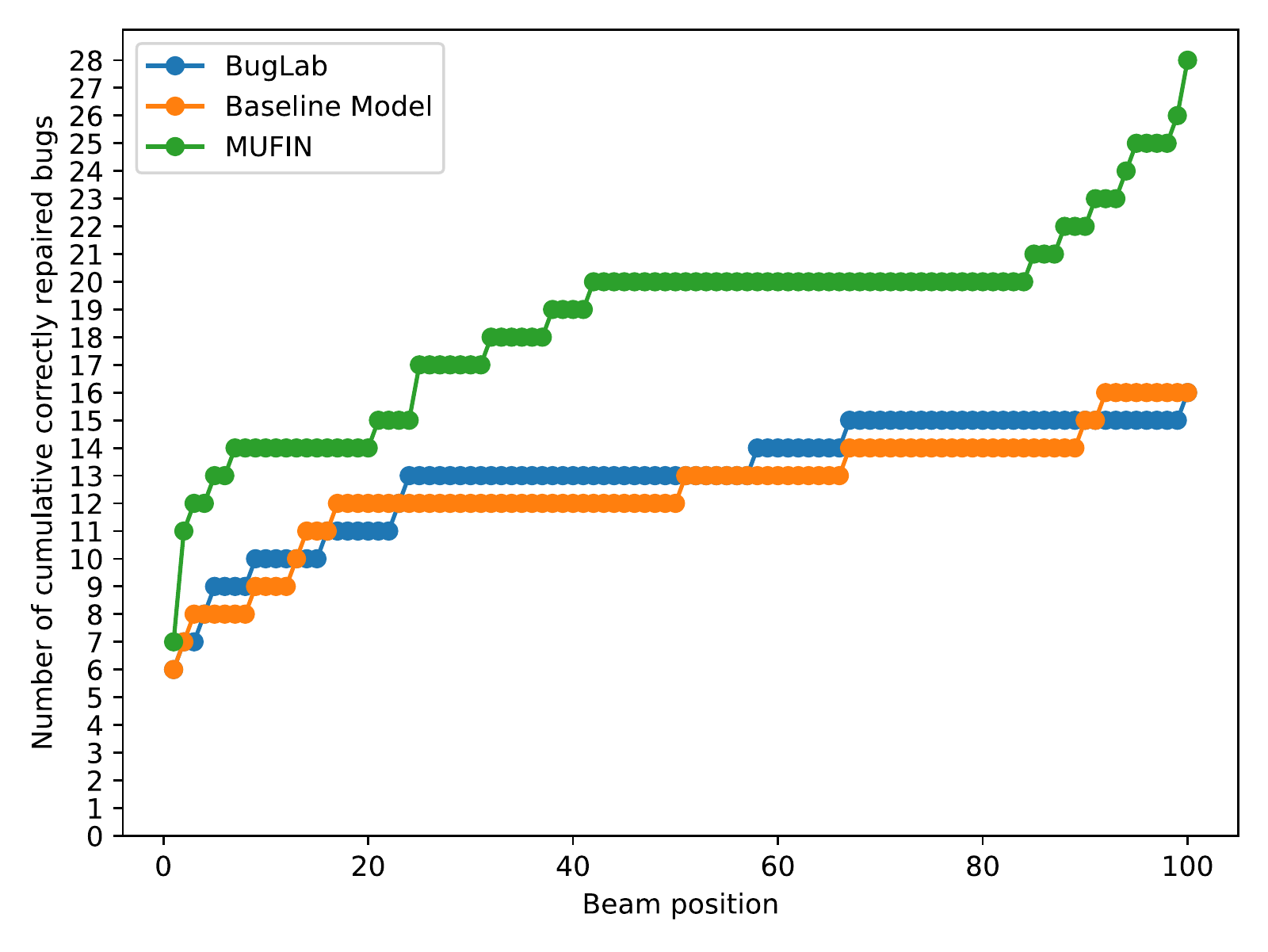}}
	\caption{Number of cumulatively correctly repaired bugs across the beam by each model on both test benchmarks. \mufin not only repairs more bugs in total but does so consistently across the beam.}
	\label{fig:repair_beam}    
\end{figure}

\paragraph{Patch Correctness over Beam}
\Cref{fig:repair_beam} plots the number of cumulatively correctly repaired bugs across the inference beam by each model on both test benchmarks.
It is useful in perceiving wherein the beam the correct patches are located.
As observed in both sub-figures per benchmark, \mufin identifies the correct patch earlier in the beam.
This indicates that \mufin improves patch prioritization inside the beam.
This means that the underlying likelihood driving the beam better captures correctness thanks to \mufin.

\paragraph{Comparison with State-of-the-Art Performance}
Per our methodology, 
we use smaller models than related work (e.g., our models have 60M parameters vs. 220M from \selfapr~\cite{ye2022selfapr}) because our goal is to precisely measure the improvement given by back-translation.
We note that those experimental results do not reflect state-of-the-art repair effectiveness \cite{jiang2023impact}.
We believe that applying our original back-translation loop to large language models would be as beneficial as what our experiments demonstrate. However, this requires computation power that is outside of our lab capacity.

\begin{tcolorbox}
    \textbf{Answer to RQ1:}
    RQ1 is based on a carefully designed protocol to measure the impact of back-translation.
    \mufin's back-translation training enables the model to correctly repair +12 (Defects4J) and +4 (QuixBugs) bugs than the baseline model.
    Since the final model comes from the same baseline, the improvement in effectiveness is causally explained by back-translation  training.
    Back-translation provides the models with more training samples and thus improves its generalization over the testing datasets.
\end{tcolorbox}

\subsection{RQ2 Results (Critics)}

In RQ2, we study the impact of the critic design in \mufin.
\Cref{tab:rq2_results} shows the effectiveness of each critic across two datasets: QuixBugs and Defects4J.
The table reads as follows.
The first meta-column gives information regarding the approach and critic used to train the model, while the second meta-column gives the number of training samples used in each stage, with column \# 3 being the number of generated samples with self-supervision in \mufin Back-Translation.
The third meta-column comprises the repair effectiveness results.
For each cell X/Y, X denotes the number of correctly repaired bugs (i.e., when the human-written tests pass), while Y indicates the number of plausibly repaired bugs (i.e., when the human-written tests pass).
The last meta-column presents the patch compilability results.
Each cell X\% denotes the percentage of compilable patches out of all generated patches for the given dataset.

\begin{table*}[t]
\centering
\caption{\mufin's effectiveness with three different critics across two testing datasets. Fine-tuning with \mufin Back-Translation always improves the fixer's effectiveness. There exists a trade-off between critic restrictiveness and effectiveness. The critic \textit{compiler} leads to the best results.}
\label{tab:rq2_results}
\begin{tabular}{c|cc|cc|cc}
\toprule
Approach & \multicolumn{2}{c|}{\# Training Samples} & \multicolumn{2}{c}{Repair} & \multicolumn{2}{c}{Patch Compilability}\\
  & Initialization & Self-Supervised &   \shortstack{QuixBugs \\ (39 bugs)} & \shortstack{D4J \\ (428 bugs)} &   \shortstack{QuixBugs \\ (39 bugs)} & \shortstack{D4J \\ (422 bugs)} \\
   Baseline Model & 3,942,935 & -                & 2/4 & 16/43          & 17.64\% & 12.73\% \\
   \mufin(\textit{no critic})  & 3,942,935 & 679,140 & 5/7 & 24/57          & \textbf{21.33\%} & \textbf{17.07\%} \\
   \mufin(\textit{compiler}) & 3,942,935   & 197,959 & \textbf{6/7} & \textbf{28/62} & \textbf{21.33\%} & 17.00\% \\   
   \mufin(\textit{tests}) & 3,942,935      & 21,196  & 5/8 & 23/51          & 16.54\% & 13.32\% \\
\bottomrule
\end{tabular}
\end{table*}

\begin{figure*}[!bt]
\begin{lstlisting}[language=diff,caption={Example of a bug generated by \mufin(\textit{compiler}) breaker. Here, an entirely new statement is synthesized and introduced by the breaker model, modifying the intended behavior of the program.},numbers=none,label=lst:breaker_compiler_perturbation]
--- a/correct.java
+++ b/buggy.java
  if (variableType == VARIABLE_TASK) {
+  response.setUrl(urlBuilder.buildUrl(RestUrls.URL_TASK, task.getId()));
   restVar.setValueUrl(urlBuilder.buildUrl(RestUrls.URL_TASK_VARIABLE_DATA, id, name));
\end{lstlisting}
\end{figure*}

\paragraph{Repair Effectiveness}
In total, all \mufin models show higher repair effectiveness than the baseline model.
\mufin models correctly repair +8 (\textit{no critic}), +12 (critic = \textit{compiler}), and +7 (critic = \textit{tests}) Defects4J bugs than the baseline model, while also finding plausible patches for a higher number of bugs.
The same is observed on QuixBugs, where \mufin models correctly repair +3 (\textit{no critic}), +4 (critic = \textit{compiler}), and +3 (critic = \textit{tests}) bugs than the initial model.
Regardless of the critic, the fine-tuning process in \mufin Back-Translation is useful. 
This is clearly evidenced that even with no critic, performance improves.

At the same time, we observe that the critic has an impact on the final repair effectiveness of the model.
The most effective model is trained using the \textit{compiler} critic, correctly repairing a total of 28 Defects4J bugs.

Let us now look at the training sample reduction due to critics.
As observed in \Cref{tab:rq2_results}, \mufin with no critic generates 679,140 self-supervised samples for fine-tuning.
The \textit{compiler} critic keeps only one-third of the data, resulting in 197,959 training samples.
Finally, we see that the \textit{tests} critic is extremely selective, resulting in only 21,196 training samples.
Since \mufin with the \textit{tests} critic performs worse than with no critic, it means that the filtering is too extreme, and the resulting scarcity of self-supervised samples is ineffective.
This is contrary to our initial expectations given that the \textit{tests} critic results in high-quality data: a bug is 100\% sure a bug and a fix is most likely a fix given that we consider projects with strong test suites.

We note that despite more permissive critics such as \textit{compiler}, nothing prevents the breaker from generating interesting bugs that are valuable for fine-tuning the models.
The breaker trained with the \textit{compiler} critic does not simply introduce trivial bugs.
For instance, consider the example of a bug introduced by \mufin(\textit{compiler})'s breaker in \autoref{lst:breaker_compiler_perturbation}.
In this case, the buggy patch introduces an entirely new statement composed of a method call that is not available in the input's context.
This bug compiles.
Even without being formally validated by a failing test case, this bug can be considered a functional bug as it introduces unintended behavior in the program.

Overall, there is a clear trade-off between critic restrictiveness (i.e., how restrictive a critic is in enforcing sample quality) and the number of training samples generated for back-translation. 
The \textit{compiler} critic is the best in that trade-off.

\paragraph{Patch Compilability}
The last meta-column of \autoref{tab:rq2_results}
shows that all \mufin models exhibit higher patch compilability rates than the initial baseline model.
For example, both \mufin(\textit{compiler}) and \mufin(\textit{nocritic}) produce 17\% of compilable patches over all bugs and a beam size of 100.
Clearly, \mufin(\textit{tests}) as the critic, despite improving repair effectiveness results w.r.t. to the baseline model, fails to do the same in patch compilability. In the case of QuixBugs, \mufin(\textit{tests})'s patch compilability rate is even lower than the baseline.

These results further emphasize the trade-off between critic restrictiveness, as a proxy of sample quality, and the number of samples available for fine-tuning.
Both \mufin(\textit{no critic}) and \mufin(\textit{compiler}) generate a higher number of samples that improve the syntactic quality of the generated patches.
The same argument explains why no critic slightly outperforms in terms of compilability, because it yields 3x more training samples.

\paragraph{Critic Execution Time}
Different critics require different amounts of time to filter samples.
We note that using the \textit{tests} critic is far more computationally intensive than using the \textit{compiler} critic, as the former not only compiles but also executes test suites.
To overcome this problem, future work might favor purely static critics that are lightweight and fast.

\begin{tcolorbox}
    \textbf{Answer to RQ2:}
    Our results show the relevance of critic design for the effectiveness of the final fixer model.
    The best critic for \mufin is the \textit{compiler} critic, which filters the top-quality one-third of the generated data, resulting in a neural fixer capable of correctly repairing 28 Defects4J bugs, 12 more bugs than the non-fine-tuned baseline.
    We clearly observed and discussed in detail the trade-off between critic restrictiveness and the quantity of available training samples.
\end{tcolorbox}

\section{Discussion}

\subsection{Breaker Repurposing}
\label{sec:breakerrepurpose}

While the goal of this work is to repair buggy programs, bugs are a central component of several other software engineering tasks.
Many approaches to such tasks require large amounts of executable bugs to either be built or executed.
For example, experimental work on fault localization \cite{li2021fault} and bug detection \cite{pradel2018deepbugs} require large amounts of bugs for achieving sound results.

Collecting such large amounts of executable bugs does not scale as it requires intense human effort.
A neural model capable of generating bugs is, therefore, one possible solution to this issue: by applying a breaker model to several locations in a few correct projects we can obtain a very large number of executable bugs.
To that extent, \mufin's breaker model has inherent value.
By design, \mufin's breaker model is trained to generate quality bugs according to the critics.

The prolificacy of \mufin's breaker to generate bugs is evidenced by our experimental results.
With the most liberal setup, \mufin(\textit{no critic}), the breaker alone generates 678,540 bugs.
That represents an average of approx. 12,117 bugs seeded per project.
Furthermore, when using a critic that retains only those samples which successfully compile and have failing unit tests (\mufin(\textit{tests}), we obtain bugs which have the guarantee to be built and have at least one failing unit test.
Our data is composed of 21,180 such bugs, which is an order of magnitude higher than the number of executable bugs available in manually curated datasets from the literature.

\begin{figure}
\begin{lstlisting}[language=diff,caption={Example of a bug generated by \mufin(\textit{tests}) breaker. The breaker closes an output stream before intended, resulting in an IOException being thrown when a sub-sequent flush operation is called.},numbers=none,label=lst:breaker_perturbation]
--- a/correct.java
+++ b/buggy.java
@@ -866,12 +866,11 @@
  _flushBuffer();
  if (flushStream) {
+  out.close();  
   _out.flush();
  }
\end{lstlisting}
\end{figure}

\autoref{lst:breaker_perturbation} shows an example of a bug generated by a \mufin breaker model.
Here, the breaker model introduces a functional bug that triggers at least one failing unit test.
More specifically, the breaker model closes an output stream before intended, releasing all system resources associated with it, by introducing a \textit{close()} method call right before a \textit{flush()} method call.
When the \textit{flush()} method call is executed, an \textit{IOException} is thrown due to the stream being closed.

Also, one can consider mutation testing \cite{jia2010analysis} as dependant on a breaker.
To that extent, the breaker could be used as a plug-and-play component in a mutation testing infrastructure, in the spirit of related work on neural mutation \cite{tufano2019learning, tufano2020deepmutation}.

For all these reasons, \mufin's usefulness outreaches program repair, and the breaker which is initially a by-product can become a valuable asset for future research.
We make our breaker models and generated bugs available to the scientific community. All bugs have the property of being  buildable and runnable with at least one failing test for experiments in automated software engineering.

\subsection{Threats to Validity}

We identify as internal threats to the validity of our results potential bugs in our implementation and errors in our manual patch analysis.
To address these, we make our implementation and experimental artifacts publicly available.

As external threat, we identify the focus on a single programming language (Java).
To mitigate external threats, we evaluate on two well-established Java program repair benchmarks.
In principle, our experimental results should generalize to arbitrary programming languages and repair benchmarks.

\subsection{Future Work}

In RQ2, we have studied the impact of the critic design on the effectiveness of the fixer models.
However, some relevant questions remain unanswered.
First, it remains unknown the extent to which the critic design depends on the effectiveness of the initial fixer/breaker models.
Second, it also remains unanswered how the seed datasets in \mufin Back-Translation impact the generalization of the fixer model.
Our intuition is that the larger, but also more diverse and realistic, the seed datasets are, the higher the impact of fine-tuning with back-translation.
Third, it remains to be studied if and how further iterations of \mufin Back-Translation improve the neural fixer model, as well as when and how this process converges.

\section{Related Work}
 
\subsection{Automated Program Repair}

\textit{Heuristic-based} approaches~\cite{forrest2009genetic,kim2013automatic,liu2019tbar,cornu2015npefix} generate and validate patch candidates by first computing the search space of modifications based on the programmed heuristics, and then by running the tentatively patched program against the set of provided tests.
The patch candidates that make the modified program pass all tests are considered correct.

\textit{Constraint-based} approaches~\cite{nguyen2013semfix,xuan2016nopol,mechtaev2016angelix} follow a different strategy.
First, they identify constraints that must be met in order to repair the bug.
Then, a program synthesis technique is guided by the identified constraints to generate patches.

\textit{Learning-based} approaches~\cite{gupta2017deepfix,chen2019sequencer,lutellier2020coconut,ye2021neural,chen2021neural,yasunaga2021break,allamanis2021self,yasunaga2020graph, ye2022selfapr} leverage pairs of buggy and correct samples to learn models which can generate patches.
Typically, deep learning techniques are employed to obtain the patch generation models.
\mufin differs from traditional supervised learning repair approaches~\cite{gupta2017deepfix,chen2019sequencer,lutellier2020coconut,jiang2021cure,ye2021neural,chen2021neural} in being self-supervised.

\subsection{Self-Supervised Learning on Code}

Self-Supervised Learning is a paradigm of machine learning which consists in transforming unpaired data into paired data to train machine learning models in a supervised manner.
It is a solution to the cost of obtaining supervised training data.
In recent years, Self-Supervised Learning has become increasingly popular in the \acf{NLP} world and has been used successfully in the task of learning word representations~\cite{mikolov2013efficient,pennington2014glove}, in pre-training language models~\cite{taylor1953cloze,devlin2018bert,liu2019roberta} and in pre-training code language models~\cite{ahmad-etal-2021-unified, wang2021codet5, feng2020codebert}.

Self-Supervised Learning has been little applied to coding tasks.
Kommrusch et al. \cite{kommrusch2021self} learn to prove program equivalence by self-supervision over complete and incomplete proofs in an iterative training procedure.
Ni et al.~\cite{nilearning} use a similar approach to generate and sample correct and partially-correct code solutions for math reasoning problems.

Self-Supervised Learning can be applied to \ac{APR}.
Loriot et al.~\cite{loriot2022styler} learn to fix Checkstyle violations by training models on artificially injected violations.
Yasunaga et al.~\cite{yasunaga2020graph} propose a self-supervised procedure for compilation bugs based on artificial bugs.
Both techniques' corruption procedures work at the character and token levels and do not aim at introducing functional bugs, a key difference from \mufin's goal.
Vasic et al.~\cite{vasic2019neural} synthetically replace variable names to generate a training dataset for a joint localize and repair model for variable misuse bugs.
Allamanis et al.~\cite{allamanis2021self} train a bug detection and repair model with artificially generated bugs of four different categories.
Ye et al.~\cite{ye2022selfapr} propose \selfapr: a self-supervised framework to train neural models with execution diagnostic information.
The key difference between these works is that none of them use back-translation, as \mufin does.

\subsection{Back-Translation on Code}

Back-translation is a Self-Supervised Learning technique for neural machine translation that generates synthetic training data by translating a text from a source language to a target language and back to the source language~\cite{edunov2018understanding, lample2018phrase}.
Rozière et al.~\cite{roziere2020unsupervised,roziere2021leveraging} apply back-translation to the task of code translation, which is not program repair.
Wang et al.~\cite{wang2022leveraging} use back-translation in their data augmentation strategy to discover causal relations between the input source (code and comments) and corrected bugs. Their goal is to improve the prediction interpretability of APR models.
Ahmad et al.~\cite{ahmad2022summarize} also use back-translation in the task of code translation with a target-to-NL-to-source objective instead of a target-to-source-to-target objective.

Drain et al.~\cite{drain2021deepdebug} use back-translation to train a repair model.
The main differences between this work and \mufin are that they do not use critics and do not consider complex functional bugs in Java.

Yasunaga and Liang~\cite{yasunaga2021break} propose \acf{BIFI} for repairing compilation errors, augmenting it by introducing critics.
The main differences between BIFI and \mufin are the following.
First and foremost, BIFI focuses on simple compilation bugs, where the patches mostly consist of single character changes, such as a missing semi-column. 
On the contrary, \mufin repairs functional bugs with failing test suites, where patches change multiple tokens and change the AST structure. 
Our experiment is the first-ever proof of concept that back-translation can be used beyond few-character changes.

\section{Conclusion}

In this paper, we presented \mufin, a novel self-supervised approach for automated program repair in Java.
\mufin introduces a novel back-translation loop with critics dedicated to functional bugs.
To demonstrate the usefulness of \mufin, we experiment across two widely accepted Java program repair benchmarks.
Our results show that \mufin improves the patch quality w.r.t. baseline models, both in terms of repair and patch compilability.
Also, we demonstrate that not all critics are equal.
The \textit{compiler} critic based on the compilation results, leads to the best and most consistent results, being able to provide neural networks with high-value training samples while not being too restrictive so that a good number of self-supervised training samples can be employed in the back-translation loop.

We note that self-supervised neural program repair is a relatively unexplored research direction. Yet, our results show that this paradigm has great potential for improving any model. To this end, we make our implementation and experimental artifacts publicly available to foster future research on this topic.

\balance
\bibliographystyle{IEEEtran}
\bibliography{IEEEabrv,main}

\end{document}